\documentclass[12pt]{iopart}
\usepackage{graphicx}
\usepackage{
amssymb,amsbsy}
\usepackage[mathscr]{eucal}
\begin{document}

\title{Cosmological extrapolation of MOND}
\author{V.V.Kiselev$^{1,2}$ and 
S.A.Timofeev$^{1,2}$}
\address{$^1$Russian State Research Center Institute for High Energy Physics,\\
Pobeda 1, Protvino, Moscow Region, 142281, Russia}
\address{$^2$Department of Theoretical Physics,\\
Moscow Institute of Physics and Technology (State University),\\
Institutsky 9, Dolgoprudny, Moscow Region, 141701,  Russia}
\eads{\mailto{Valery.Kiselev@ihep.ru}, 
\mailto{serg\_timofeev@list.ru}}

\begin{abstract}{Regime of MOND, which is used in astronomy to describe the
gravitating systems of island type without the need to postulate the
existence of a hypothetical dark matter, is generalized to the case of
homogeneous distribution of usual matter by introducing a linear dependence
of the critical acceleration on the size of region under consideration. We
show that such the extrapolation of MOND in cosmology is consistent with both
the observed dependence of brightness on the redshift for type Ia supernovae
and the parameters of large-scale structure of Universe in the evolution,
that is determined by the presence of a cosmological constant, the ordinary
matter of baryons and electrons as well as the photon and neutrino radiation
without any dark matter.}
\end{abstract}

 \pacs{98.80.-k, 04.50.Kd}

\maketitle

\section{Introduction}

\begin{figure}[b]
\setlength{\unitlength}{1mm}
\begin{center}
\begin{picture}(50,77)
\put(5.5,3.5){  \includegraphics[width=51\unitlength]{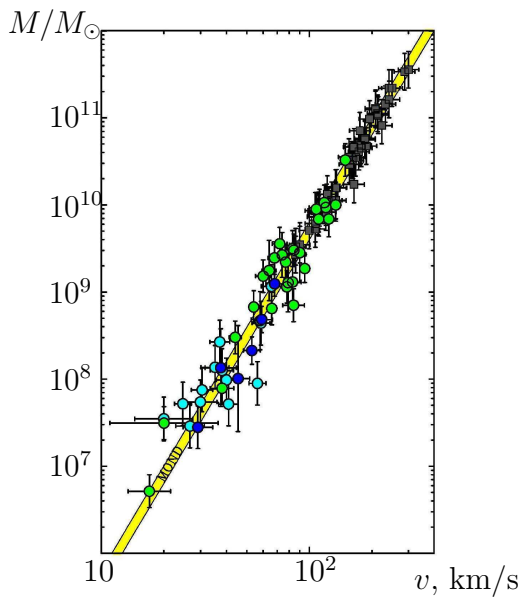}}
\put(52,2){$v$, km/s}
\put(8,3){$10$}
\put(36,3){$10^2$}
\put(3.4,17){$10^7$}
\put(3.4,28.5){$10^8$}
\put(3.4,40){$10^9$}
\put(3.4,51.5){$10^{10}$}
\put(3.4,63.5){$10^{11}$}
\put(-2,76){$M/M_\odot$}
\end{picture}
\end{center}
  \caption{The correlation of visible baryonic masses in the disk galaxies with
the rotation velocities of stars in the region of flat rotation curves,
as shown in comparison with the MOND prediction: $v^4=GM\tilde g_0$.
The mass is given in units of solar mass $M_\odot$. The figure is
taken from \cite{rev-MOND1}.}\label{fig-TF}
\end{figure}

In the framework of modified Newtonian dynamics (MOND) \cite{MOND} we get
quite the reasonable theoretically and successful phenomenologically
explanation for the empirical Tully--Fisher law, that regularly relates the
\textit{visible} masses of spiral galaxies to asymptotically constant
rotation-velocities of stars in the region of dominating ``dark matter halo''
(see figure \ref{fig-TF}) as well as for profiles of rotational velocities of
stars in galaxies, wherein the regime of flattening the rotation curves does
not take place. Moreover, the MOND is able to explain many other phenomena
caused by the \textit{inhomogeneity} of baryonic matter-distribution in the
Universe at the scales of gravitationally coupled systems (see reviews in
\cite{rev-MOND1,rev-MOND2,rev-MOND3}).
For instance, in disk galaxies, wherein a dominant contribution to the
baryonic mass is given by the interstellar gas, the matter density is
determined with a high accuracy, since an uncertainty, caused by the
extraction of stellar masses from their visible magnitudes, is significantly
reduced. The MOND gives a precise agreement of its predictions with the
experimental data on the rotation speeds of matter around the centers of
galaxies without any free parameters, while the uncertainties of results are
caused by errors of measurements, only \cite{McGaugh}.

By construction, the basic conclusion of MOND is the statement that the
observational data can be confidently described without any introduction of
non-baryonic dark matter at the galactic scales of matter-inhomogeneity as
well as at the scales of gravitationally coupled galactic clusters
\cite{rev-MOND2} due to the specially fitted universal modification of the
gravity law if the acceleration of free falling $g$ is less than the critical
value $\tilde{g}_0\approx 1.2\cdot 10^{-10}\mbox{m/s}^2$, so that
$g^2=GM\tilde g_0/r^2$, where $M$ is the matter mass, while $r$ is a distance
to a reference point.

However, such the conclusion conflicts with cosmological modeling the
Universe evolution and properties of its large scale structure, that is
characterized by a high degree of homogeneity for the spatial distribution of
matter ($\delta \rho/\rho\sim \delta T/T\sim 10^{-5}$ for fluctuations of
energy density $\rho$ and temperature of cosmic microwave background
radiation (CMBR), $T$), because such calculations lead to the necessary
introduction of dark matter with the density being approximately 5 times
greater than the baryonic density \cite{WMAP}, if we follow the equations of
general relativity (GR).

In cosmology of homogeneous and isotropic Universe, the points in the space
with coordinates $\boldsymbol r$ move as given by a dependence of scale
factor on the time, $a=a(t)$, so that velocity $\boldsymbol v$ and
acceleration $\boldsymbol g$ of material point are determined by
\begin{equation}\label{move}
    \boldsymbol x=a(t)\boldsymbol r,\quad
    \boldsymbol v=\dot{\boldsymbol x}=\dot a\, \boldsymbol r,\quad
    \boldsymbol g=\ddot{\boldsymbol x}=\ddot a\,\boldsymbol r,
\end{equation}
where the dot over a symbol denotes the derivative with respect to time. Due
to the homogeneity and isotropy, the evolution equations do not contain the
co-moving coordinate $\boldsymbol r$. Therefore, the evolution law is
universal, and it can be considered at small $\boldsymbol r$, i.e. in the
region of applicability of the non-relativistic mechanics ($\boldsymbol v\to
0$) and gravity law by Newton, namely, the acceleration of free falling along
the radius vector $g=\ddot a\, r$ reads off
\begin{equation}\label{force}
    g=-\frac{G\mathcal{M}_{KT}}{|\boldsymbol x|^2},
\end{equation}
where we introduce the Komar--Tolman gravitational mass\footnote{The source
of gravitational field in Einstein's equations is the tensor of
$2(T_{\mu\nu}-\frac12 T\,g_{\mu\nu})$, where the symbol $T$ corresponds to
the tensor of energy-momentum and its trace, and $g_{\mu\nu}$ denotes the
metric. Therefore, the static gravitational potential is determined by not
the energy density alone, but by the temporal component of the source. The
integration of temporal component over the volume gives the Komar--Tolman
mass, exactly.} for a spherically symmetric distribution of the matter with
the energy density $\rho$ and pressure $p$ inside the sphere with radius
$|\boldsymbol x|$:
\begin{equation}\label{K-T}
    \mathcal{M}_{KT}=\frac{4\pi}{3}\,(\rho+3p)\,|\boldsymbol x|^3,
\end{equation}
so that
\begin{equation}\label{ddot}
    \frac{\ddot a}{a}=-\frac{4\pi G}{3}\,(\rho+3p),
\end{equation}
that naturally does not include the co-moving coordinate $r$.

In addition, the adiabatic law of energy conservation for the substance with
the energy density $\rho$ and pressure $p$ reads off
\begin{equation}\label{conserv}
    d\mathcal{E}=-pd\mathcal{V}
\end{equation}
so under the substitutions of $\mathcal{E}=\rho \mathcal{V}(t)$,
$\mathcal{V}(t)=a^3\mathcal{V}_0$, we get
\begin{equation}\label{conserv2}
    \dot \rho+3H(\rho+p)=0.
\end{equation}
Then, after multiplying (\ref{ddot}) by $2\dot a a$ we can easily integrate
and get the Friedmann equation
\begin{equation}\label{Hubble1}
    H^2=\frac{8\pi G}{3}\left(\rho+\frac{\rho_E}{a^2}\right),
\end{equation}
where we have introduced the Hubble constant $H=\dot a/a$, while in GR the
constant of integration $\rho_E$ is presented by the sum of contributions
given by a constant spatial curvature and a matter with equation of state
$p=-\frac13\,\rho$. These terms do not contribute to the Komar--Tolman mass,
i.e. they do not produce the gravitational force.

Thus, we get the equations of GR for the evolution of homogeneous and
isotropic Universe (\ref{conserv2}) and (\ref{Hubble1}). Then, the
description of such evolution is reduced to the non-relativistic dynamics at
$r\to 0$.

The straightforward application of MOND paradigm at $r\to 0$ implies the
transition to the regime of modified law of gravity, when
$$
    g\mapsto g\left|\frac{g}{\tilde{g}_0}\right|,\qquad\mbox{at }
    |g|\ll \tilde{g}_0,
$$
hence,
\begin{equation}\label{mond-old}
    \frac{\ddot a}{a}\,|\ddot a|=-\frac{4\pi G}{3}\,(\rho+3p)\,\frac{\tilde{g}_0}{r}.
\end{equation}
Notice that a direct observation of the scale factor evolution by the
registration of type Ia supernovae \cite{Riess,SN1,SN2,SN3,SN4} reliably
shows that the accelerated Universe expansion ($\ddot a>0$) at present has
changed the decelerated expansion ($\ddot a<0$), which took place at the
redshift\footnote{The redshift is related to the scale factor by
$z=\frac{1}{a(t)}-1$.} $z>z_t$, where $z_t\approx 0.4-1.0$ is extracted in
the procedure of fitting the deceleration parameter $q(z)=-\ddot a/(a H^2)$
by the linear function $q(z)\mapsto q_0+z\,q_0^\prime$. Consequently, in
vicinity of transition point $z_t$, where $\ddot a=0$, the validity of MOND
in cosmology is quite justified.

However, eq. (\ref{mond-old}) at $\tilde{g}_0=\mbox{const}$ is inconsistent
with the condition of homogeneity and isotropy of matter, since the actor
$\frac1r$ explicitly depends on the co-moving coordinate $r$, hence, such the
modification of gravity law is inherently related to the inhomogeneity of
matter distribution. In particular, eq. (\ref{mond-old}) results in the
\textit{vacuum catastrophe}: the vacuum with $\rho_\Lambda=
-p_\Lambda=\mbox{const}$ is to be unstable, namely, due to eq.
(\ref{mond-old}) the inhomogeneous distribution of matter is initiated. Thus,
the straightforward transfer of MOND to the cosmology is theoretically
inconsistent. Nevertheless, keeping in mind that MOND is related to the
spatial inhomogeneity of matter distribution, we can easily suggest its
cosmological extrapolation for the homogeneous and isotropic Universe,
indeed: by setting
\begin{equation}\label{g0}
    \tilde{g}_0\mapsto g_0={g}_0^\prime |\boldsymbol x|,
\end{equation}
we arrive to the evolution equation at $|\ddot a/a|\ll g_0^\prime$
\begin{equation}\label{mond-new}
    \frac{\ddot a}{a}\,\frac{|\ddot a|}{a}=-\frac{4\pi G}{3}\,(\rho+3p)\,g_0^\prime,
\end{equation}
which is consistent with the initial conditions imposed on the matter as well
as with the vacuum stability.

On the other hand, the visible large scale structure (LSS) is seen at the
angle distances of $\delta\theta_\mathrm{LSS}\approx 1.5-2^\circ\sim
\frac{1}{40}-\frac{1}{30}$ radians, that corresponds to $|\boldsymbol
x|_\mathrm{LSS}\sim \delta\theta_\mathrm{LSS}\cdot x_H$, where the Hubble
horizon $x_H\approx \frac{1}{H_0}$ at the current value \cite{WMAP}
$$H_0\approx (69-72) \frac{\mbox{\small km}}{\mbox{\small s}\cdot\mbox{\small Mpc}}.$$
Empirically in MOND \cite{rev-MOND1}
\begin{equation}\label{empirMOND}
    \tilde{g}_0\approx\frac{H_0}{2\pi},
\end{equation}
so that if we put
$$
    \tilde{g}_0\approx g_0^\prime|\boldsymbol x|_\mathrm{LSS},
$$
then we expect
\begin{equation}\label{empir-g0}
    g_0^\prime\approx\frac{1}{2\pi\delta\theta_\mathrm{LSS}}\,H_0^2.
\end{equation}
It is convenient to parameterize the quantity $g_0^\prime$ in the form
\begin{equation}\label{K}
    g_0^\prime=K_0H_0^2,
\end{equation}
whereas by the order of magnitude
\begin{equation}\label{estimateK}
        K_0\sim\frac{1}{2\pi\delta\theta_\mathrm{LSS}}\sim 10.
\end{equation}

Thus, we can assume that the change of regime in the cosmological
extrapolation of MOND as given by (\ref{g0}) is caused by the presence of
spatial inhomogeneity of matter, that allows us theoretically to conform the
existence of two regimes for the critical acceleration of gravity.

In this paper we investigate the scheme proposed in (\ref{g0}) and (\ref{K})
for the cosmological extrapolation of MOND with the parametrization in the
transition region as derived in the holographic description of gravity as the
entropic force \cite{Verlinde,TPad} at low temperatures (the MOND regime)
\cite{KT-MOND}. This way essentially differs from the relativistic theory of
gravity for the MOND paradigm offered by J.Bekenstein in \cite{TVS}, wherein
he constructed the action satisfying a set of requirements, which include the
appropriate limit to the MOND regime as well as, in principle, a correct
description of cosmological effects, too. The modification is very specific
due to additional gravitational scalar and vector fields in the action.
Additional fields appear in the modified gravity formulated by J.W.Moffat in
\cite{Moffat}, too. Then, the gravitational constant becomes the field
satisfying the equations of motion. So, J.W.Moffat has obtained the
modification of gravity law \cite{Moffat2} similar to the MOND, but it is
distinguishable from the MOND, while some cosmological effects can be treated
in agreement with observations \cite{MT}. Thus, approaches in \cite{TVS} and
\cite{Moffat} represent the axiomatic attempts to construct the consistent
theory of modified gravity from the primary principles of field theory, and
they are able to produce the framework for complete calculations including
the propagation of cosmological perturbations in the relativistic way, of
course. We follow a more pragmatic way: without an introduction of any new
additional notions, the modification of gravitational dynamics in the form of
(\ref{g0}) and (\ref{K}) can be consistent with the MOND at galactic scales,
and it is enough for the investigation of modified cosmological evolution
essentially different from the evolution in the framework of general
relativity if we consider the ordinary components of matter. Therefore, the
extrapolated MOND could exhibit some general features, which are common for
meaningful extensions of general relativity as they tend to exclude the dark
matter from the cosmology \textit{dynamically}. In this way, we understand
that the specific additional gravitational degrees of freedom defined
explicitly as in \cite{TVS} and \cite{Moffat} should be inevitably considered
if we try to take into account the modification of theory for the propagation
of matter inhomogeneity in cosmology, hence, in the construction of full
range description of large scale structure of Universe, that includes the
matter power spectrum and anisotropy of cosmic microwave background
radiation. In the general relativity, that is the dark matter, which is
responsible for the actual description of all features in the large scale
structure due to corrections to the matter distributions during the
cosmological evolution, but it is commonly known that basic cosmological
characteristics such as the accelerated-decelerated regimes of Universe
expansion, and the visible angular scale of matter inhomogeneity in the sky
are mainly given by the appropriate evolution rate of Universe, and we show
that such the evolution can be achieved in the framework of cosmological
extrapolation of MOND without any dark matter. This extrapolation does not
provide us with the complete theory of inhomogeneity propagation (especially,
because such the theory is expected to be nonlinear in contrast to the
well-working ordinary linear perturbation theory in the general relativity),
hence, we do not suppose to reproduce the full description of angular
dependence of cosmic microwave background radiation subject to the given
primary spectrum of matter distribution, of course, or the big-bang
nucleosynthesis, for instance. Nevertheless, the baryonic matter
inhomogeneities themselves can produce the potential gravitational wells for
the further concentration of matter, though such the wells would be affected
by the sound propagation in the photon-electron-baryon plasma, and this
evolution of inhomogeneities will be mainly determined by the strong field
regime of MOND, which is consistent with the general relativity, while the
nonlinear effects are expected to be subleading. Thus, we can expect the
correct reproduction of main features of large scale structure in the
framework of cosmological extrapolation of MOND. So, our preliminary studies
show that the first acoustic peak in the cosmic microwave background
anisotropy is successfully obtained within the cosmological extrapolation of
MOND under the very standard settings of primary spectrum of matter power
without any dark matter, of course, while the second peak can be also
correctly described by an appropriate variation of the matter power spectrum,
but the subleading structures need the construction of full theory of cosmic
perturbations beyond of purposes of our paper. So, the analysis of anisotropy
in the cosmic microwave background radiation would be presented elsewhere,
since it requires a deep and comprehensive investigation in details. The
similar note can be addressed to the problem of nucleosynthesis, too. Here,
we present the general scheme for the cosmological extrapolation of MOND and
its applications to the basic features of Universe evolution.

In this paper, we apply the minimal scheme of (\ref{g0}) and (\ref{K}) to the
description of data on type Ia-supernovae. We show that in the redshift
region accessible for observations, such the astronomical data can be
reliably described in the proposed model if there are the cosmological
constant and baryonic matter, only, without any dark matter, whereas the
numerical estimate of (\ref{estimateK}) is correct. While comparing with the
Friedmann cosmology at presence of dark matter, we can evaluate the imitated
contribution of dark matter $\rho_M$ to the critical density of
$\rho_c=3H_0^2/8\pi G$. The ratio of $\Omega_M=\rho_M/\rho_c$ to the baryonic
contribution $\Omega_b=\rho_b/\rho_c$ is roughly determined by $K_0$ in order
of magnitude. In accordance to the standard Friedmann cosmology-model with
the cold dark matter and cosmological constant ($\Lambda$CDM), the parameters
correspond to $\Omega_M\approx 0.27$ at $\Omega_b=0.045$.

Further, we analyze the large scale structure of inhomogeneity in the matter
distribution of Universe in calculations of baryon acoustic oscillations,
measured by observing the luminous matter in the sky versus the redshift
\cite{BAO}, and in estimation of ``acoustic scale'' in the CMBR anisotropy
\cite{WMAP}. The estimates obtained in the framework of cosmological
extrapolation of MOND agree with the measured values of LSS without any
postulation of hypothetic dark matter: it is enough to adjust the value of
baryonic matter in the presence of cosmological constant. Then, the density
of baryonic matter is roughly twice its value in $\Lambda$CDM of GR.

After modeling the acoustic scale in the CMBR anisotropy we calculate the
critical acceleration $\tilde g_0$ in the framework of the cosmological
extrapolation of MOND and find preferable values of cosmological parameters
of the Universe (the density of baryonic matter). We analyze the evolution
into the future, too.

In Conclusion we summarize our results and discuss its meaning.

\section{MOND parameters from data on supernovae}

At present, measuring the stellar magnitude of type Ia supernovae versus the
redshift gives the only direct observation of scale factor versus time in
cosmology. Such the observation does not base on any model of Universe
expansion, so that even the theory beyond the Friedmann equation can be
confronted to the experiment without any additional assumptions. In this
respect, such the data are unique and they are suitable for testing the
cosmological extrapolation of MOND.

The stellar magnitude $\mu$ depends on the redshift according to the formula
\begin{equation}\label{mu}
   \mu=\mu_{abs}+5\log_{10}d_L(z) + 25,
\end{equation}
where $\mu_{abs}$ is the absolute stellar magnitude, i.e. the stellar
magnitude of light source at the distance of 10 pc. The photometric distance
$d_L$ measured in Mpc, is determined by the Hubble constant evolution
\begin{equation}\label{dl}
    d_L(z)=(1+z)\int\limits_0^z\frac{c\,dz}{H(z)},
\end{equation}
where $c$ is the speed of light. In \cite{Riess} the following dependence of
the deceleration parameter on the redshift has been used as the working
hypothesis:
\begin{equation}\label{q-lin}
    q(z)\mapsto q_\mathrm{lin.}=q_0+zq_0^\prime,
\end{equation}
that allows it to describe the supernovae data at values
\begin{equation}\label{q-fit}
    q_0=-0.85\pm0.35,\qquad q_0^\prime=1.8\pm1,
\end{equation}
essentially different from zeros, whereas the parameters significantly
correlate (see details in the original paper \cite{Riess}).

According to (\ref{q-lin}), the Hubble constant
\begin{equation}\label{H-lin}
    H_\mathrm{lin.}=H_0
    (1+z)^{1+q_0-q_0^\prime}\exp\{q_0^\prime{z}\},
\end{equation}
that should be reasonably compared with the case of zero deceleration
parameter, when
\begin{equation}\label{H-nil}
    H_\mathrm{nil}=H_0(1+z).
\end{equation}
From eqs. (\ref{mu})--(\ref{H-nil}) we see that normalizing the data at low
redshifts, i.e. measuring the value of $H_0$, allows us to plot a fiducial
dependence of supernovae stellar magnitude $\mu_0$ with the deceleration
parameter equal to zero, and further, systematically to study the deviations
from this dependence versus the increasing redshift.

\begin{figure}[t]
\setlength{\unitlength}{0.9mm}
\begin{center}
\begin{picture}(145,108)
\put(4.2,2.7){  \includegraphics[width=143\unitlength]{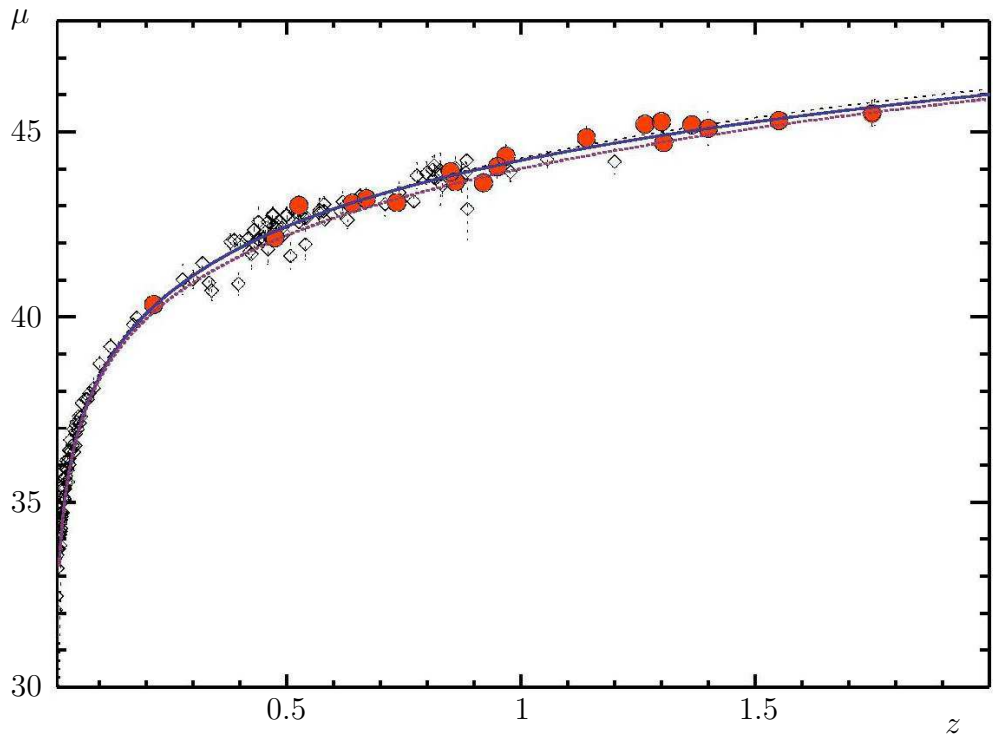}}
 \put(140,0){$z$}
 \put(39.75,2){$0.5$}
 \put(76.4,2){$1$}
 \put(109.5,2){$1.5$}
 \put(2,5.5){$30$}
 \put(2,32.7){$35$}
 \put(2,60.){$40$}
 \put(2,87.5){$45$}
 \put(2,105){$\mu$}
\end{picture}
\end{center}
\caption{The stellar magnitude $\mu$ of type Ia supernovae versus the
redshift $z$: the bottom dotted curve presents the case of zero deceleration
parameter, the solid curve shows the result of cosmological model of MOND,
the top dotted curve depicts the fit in $\Lambda$CDM. The supernovae
discovered by the ground based telescopes are marked by rhombuses, the
circles show the supernovae discovered by the Hubble Space Telescope (data
from \cite{Riess}).
  }\label{ev1}
\end{figure}

\begin{figure}[t]
\setlength{\unitlength}{0.89mm}
\begin{center}
\begin{picture}(153,100)
\put(4.5,3.5){
\includegraphics[width=146.5\unitlength]{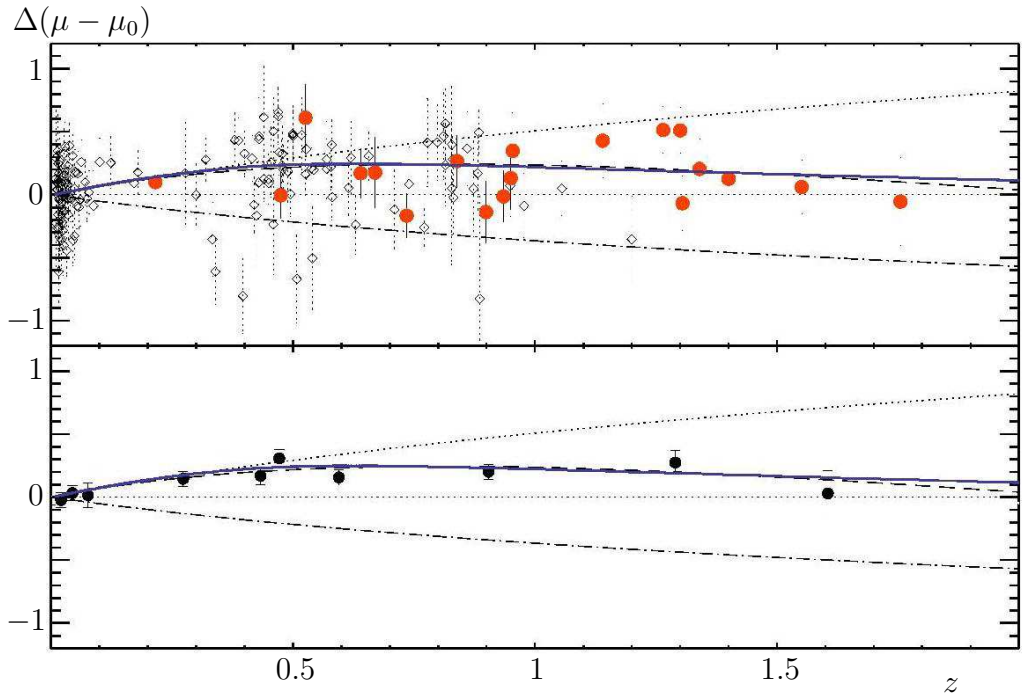}}
 \put(140,0){$z$}
 \put(40.3,2){$0.5$}
 \put(78.15,2){$1$}
 \put(112.7,2){$1.5$}
 \put(0,9.3){$-1$}
 \put(3.3,28){$0$}
 \put(3.3,46.75){$1$}
 \put(0,54.3){$-1$}
 \put(3.3,73){$0$}
 \put(3.3,91.5){$1$}
 \put(1,99){$\Delta(\mu-\mu_0)$}
\end{picture}
\end{center}
  \caption{The difference of stellar magnitudes $\Delta(\mu-\mu_0)$
  for the type Ia supernovae versus the redshift $z$ after subtracting
  the fiducial magnitude at zero deceleration parameter. The solid line
  refers to the cosmological MOND, the dashed line shows the best fit
  at the deceleration parameter $q(z)=q_0+z \,q_0^\prime$ \cite{Riess}.
  The bottom panel serves for the systematical illustration of
  supernovae data as averaged for the similar redshifts. The dotted and
  dashed-dotted lines represent the evolution with appropriate constant
  positive and negative deceleration parameter, respectively. }\label{dif-ev1}
\end{figure}

Data from \cite{Riess} are shown in figures \ref{ev1} and \ref{dif-ev1}. They
confidently signalize on the accelerated expansion of Universe at present as
well as on the existence of transition from the decelerated expansion to the
accelerated one at $z=z_t\sim 0.5$. The best fit of data takes place at
parameter values listen in (\ref{q-fit}), as it shown in figure \ref{dif-ev1}
by the dashed curve.

In the cosmological extrapolation of MOND we use the following equation for
the acceleration at the presence of both the cosmological term with the
energy density $\rho_\Lambda$ and baryonic matter with the density $\rho_b$:
\begin{equation}\label{CexMOND2}
    \frac{\ddot a}{a}\,\mathscr{D}\left(
    \frac{\pi^2}{6}\,\frac{a g_0^\prime}{|\ddot a|}
    \right)=\frac{4\pi G}{3}\left(2\rho_\Lambda-\rho_b\right)\hskip-2pt,
\end{equation}
where the region of transition from the Friedmann regime to the MOND is
described by the 1-dimensional Debye function as accepted in \cite{KT-MOND},
hence,
\begin{equation}\label{Debye}
    \mathscr{D}(x)=\frac{1}{x}\int\limits_0^x \frac{y \mathrm{d} y}{\exp\{y\}-1},
\end{equation}

\noindent

\noindent so that at the acceleration greater than critical one, i.e. in the
limit of standard cosmology at $|\ddot a|/a\gg g_0^\prime$, we get
$$
    \mathscr{D}\left(
    \frac{\pi^2}{6}\,\frac{a g_0^\prime}{|\ddot a|}
    \right)\to 1,
$$
while in the deep regime of MOND, i.e. at $|\ddot a|/a\ll g_0^\prime$,
asymptotically we find
$$
    \mathscr{D}\left(
    \frac{\pi^2}{6}\,\frac{a g_0^\prime}{|\ddot a|}
    \right)\to \frac{|\ddot a|}{a g_0^\prime},
$$
and we arrive to eq. (\ref{mond-new}).

The critical density defined by
$$
    \rho_c=\frac{3}{8\pi G}\,H_0^2,
$$
allows us to write down the baryonic density as
$$
    \rho_b=\rho_c \frac{\Omega_b}{a^3},
$$
while the requirement of vanishing the acceleration at redshift $z_t$ leads
to the following expression for the cosmological contribution:
$$
    \rho_\Lambda=\frac12\,\rho_c \Omega_b (1+z_t)^3.
$$
Thus, the model for the cosmological extrapolation of MOND is described by
equation
\begin{equation}\label{CexMOND2}
    \frac{\ddot a}{a}\,\mathscr{D}\left(
    \frac{\pi^2}{6}\,\frac{a}{|\ddot a|}\,K_0H_0^2
    \right)=\frac12\,H_0^2\Omega_b\left((1+z_t)^3-\frac{1}{a^3}\right)\hskip-3pt,
\end{equation}
combined with initial data
$$
    a_0=1,\qquad \dot a_0=H_0.
$$
Figures \ref{ev1} and \ref{dif-ev1} illustrate the results of cosmological
MOND at the typical value of $\Omega_b=0.045$ as in $\Lambda$CDM, while
\begin{equation}\label{parCexMOND}
    q_0=-0.775,\qquad z_t=0.375,\qquad\Rightarrow\;K_0=16.7,
\end{equation}
so that in the region  of available data, the calculated profile is
practically indistinguishable from the fit obtained at the linear dependence
of deceleration parameter on the redshift as performed in \cite{Riess}. Note,
that the difference of stellar magnitudes presented in figure \ref{dif-ev1}
is independent of the normalization of Hubble constant $H_0$, while data
shown in figure \ref{ev1} allows us to fix both the value of $H_0$ at small
$z$ and the fiducial curve at $\ddot a\equiv 0$.

Notice that at present, the regime of deep MOND takes place in cosmology, so
that from eq. (\ref{CexMOND2}) we find
\begin{equation}\label{K0}
    q_0^2=\frac12\,K_0\Omega_b\left((1+z_t)^3-1\right).
\end{equation}
Hence, we can see that the value of baryonic matter density, i.e the
parameter of $\Omega_b$, can get quite arbitrary variations in this test on
the supernovae data, because the region of MOND applicability near $\ddot
a=0$ is completely described by the factor of $K_0\Omega_b$, entering in eq.
(\ref{K0}), while the Hubble constant at low redshifts is mainly determined
by initial data, namely, by $H_0$ and $q_0$, so that the influence of
$\Omega_b$ value on the quality of data description is quite weak. For
instance, the twice increase of baryonic fraction is almost invisible on the
curve derived from the cosmological MOND as shown in figure \ref{dif-ev1},
i.e. the region of low redshifts is weakly sensitive to $\Omega_b$.

In the framework of $\Lambda$CDM the deceleration parameter and redshift of
transition from the deceleration to the acceleration are given by the
fractions of densities for the vacuum $\Omega_\Lambda$ and matter $\Omega_M$,
\begin{equation}\label{stand-q0zt-1}
    q_0=\frac12\,\Omega_M-\Omega_\Lambda,\qquad
    (1+z_t)^3=2\frac{\Omega_\Lambda}{\Omega_M},
\end{equation}
so that in the case of spatially flat Universe, when
$\Omega_\Lambda+\Omega_M=1$, we get
\begin{equation}\label{stand-q0zt}
    q_0=\frac32\,\Omega_M-1,\qquad
    (1+z_t)^3=2\frac{1-\Omega_M}{\Omega_M}.
\end{equation}

The quality of fitting the supernovae data for the flat Universe in
$\Lambda$CDM at $\Omega_\Lambda\approx 0.73$, $\Omega_M\approx 0.27$ is
slightly worse than in the case of the linear approximation (\ref{q-lin}) for
the deceleration parameter, but it has got a little bit of better quality at
$\Omega_\Lambda\mapsto \tilde\Omega_\Lambda\approx0.95$, $\Omega_M\mapsto
\tilde\Omega_M\approx0.45$ \cite{Riess}.

The parameters brought from $\Lambda$CDM into MOND give
\begin{equation}\label{ratio}
    \frac{\Omega_M}{\Omega_b}=-\frac{K_0}{q_0}=\frac{K_0}{\Omega_\Lambda-\frac12\,\Omega_M}.
\end{equation}
At $|q_0|\sim 1$ we find $\Omega_M/\Omega_b\sim K_0$. However, by taking into
account the fact that the point of maximal confidence level of data fit in
$\Lambda$CDM is shifted from the standard values $\Omega_\Lambda\approx
0.73$, $\Omega_M\approx 0.27$ to $\Omega_\Lambda\mapsto
\tilde\Omega_\Lambda\approx0.95$, $\Omega_M\mapsto
\tilde\Omega_M\approx0.45$, hence, numerically
$\tilde\Omega_M/\Omega_b\approx 10$, we deduce $K_0\approx7.5$, that is in
agreement with our estimates obtained in consideration of changing the regime
of homogeneous matter distribution to the limit of inhomogeneity in MOND.

Thus, in the framework of cosmological extrapolation of MOND we can reliably
describe the type Ia supernovae data by making use of baryonic matter and
cosmological constant, only, without any postulating the dark matter. In this
way, the parameters of extrapolation is naturally conformed with the
Milgrom's critical acceleration in MOND, applicable at scales, when the
spatial inhomogeneity of baryonic matter distribution becomes significant.
The cosmological extrapolation of MOND allows us, in fact, to calculate the
present ratio of density of hypothetical dark matter to the baryonic density
$\Omega_M/\Omega_b$ in terms of the slope of critical acceleration in its
dependence on the distance, i.e. the parameter of $g_0^\prime$, expressed in
units of Hubble constant at present. Therefore, the supernovae data lead to
the necessary introduction of dark matter if we keep the Friedmann evolution,
i.e. if the evolution is determined by the Newton law of gravity, indeed,
while the modification of gravity law at low accelerations in the case of
inhomogeneous distribution of matter in accordance with MOND as well as for
the homogeneous distribution of matter in cosmological MOND, gives the
confident description of data without any dark matter.

\section{The large scale structure of Universe}

The Universe evolution essentially transforms the spatial distribution of
matter, that is observed as baryonic acoustic oscillations (BAO) \cite{BAO}
and CMBR anisotropy \cite{WMAP}.

So, the results of survey of the matter distribution on the celestial sphere
versus the redshift \cite{BAO} are particularly reduced to the ratio of sound
horizon $r_s$ for the waves in the baryon-electron-photon substance as
written in co-moving coordinates
\begin{equation}\label{rs}
    r_s(z)=\int\limits_0^{t(z)}c_s\,dt
\end{equation}
to the effective distance\footnote{We consider the flat space.} defined by
\begin{equation}\label{DV}
    D_V(z)=\left\{(1+z)^2D_A^2(z)\,\frac{c z}{H(z)}\right\}^{\frac13},
\end{equation}
where $c_s$ is the sound speed in the medium, while the angular distance is
given by
\begin{equation}\label{DA}
    D_A(z)=\frac{c}{1+z}\,\int\limits_0^z\frac{dz}{H(z)}.
\end{equation}
The measurements give the value of $r_s(z_d)/D_V(z)$ at $z=0.2$ and $z=0.35$,
which are shown in table  \ref{table}. Here $z_d$ is the redshift of epoch,
when the interaction of baryonic and photonic components of substance via the
Compton scattering of electrons off photons and Coulomb attraction of
electrons to protons becomes negligible, i.e. when baryons stop to drag
photons (the finish of drag epoch).
\begin{table}
  \caption{The comparison of observed parameters for the large scale structure
  of Universe with estimates in the framework of the cosmological extrapolation
  of MOND (the parameters are explained in the text).}
  \label{table}
  \centering
  $
  \begin{array}{lccc}
\br  &&&\\[-3mm]
    \mbox{quantity} & \displaystyle\frac{r_s(z_d)}{D_V(0.2)}  &
    \displaystyle\frac{r_s(z_d)}{D_V(0.35)}  & l_A \\[3mm]\hline
    \mbox{exp.\footnotesize\cite{WMAP,BAO}} & 0.1905\pm 0.0061 & 0.1097\pm
	0.0036 & 302.69\pm 0.76\\[1mm]
    \mbox{MOND} & 0.1924 & 0.1149 & 302.5\\ \br
  \end{array}
  $
\end{table}

The speed of sound is given by the following expression:
\begin{equation}\label{sound}
    c_s=\frac{c}{\sqrt{3}}\,\frac{1}{\sqrt{1+R}},
\end{equation}
where $R$ is the baryon-photon ratio depending on the redshift
\begin{equation}\label{R}
    R=\frac34\,\frac{\rho_b}{\rho_\gamma}=\frac34\,
    \frac{\Omega_b}{(1+z)\Omega_\gamma}.
\end{equation}
Introducing the redshift $z_{eq}$ for the moment, when the density of
non-relativistic matter $\rho_M$ equals to the density of radiation (photons
and neutrinos), allows us to write down the formula for the sound horizon as
\begin{equation}\label{rs2}
    r_s(z)=\frac{2}{k_{eq}}\,\sqrt{\frac{6}{R_{eq}}}
    \ln\frac{\sqrt{1+R}+\sqrt{R+R_{eq}}}{1+\sqrt{R_{eq}}},
\end{equation}
where
$$
    k_{eq}^2=2\Omega_MH_0^2z_{eq}.
$$
Analytical parameterizations of numerical calculations for $z_{eq}$, $z_d$
and $R$ are given in \cite{EH97}, wherein the dependencies on the CMBR
temperature $T=\Theta\cdot 2.7$ K and parameters $\Omega_bh^2$, $\Omega_Mh^2$
with $h$ being the coefficient for the Hubble constant written down as
$H_0=h\cdot 100\, \mbox{km}\cdot\mbox{s}^{-1}/\mbox{Mpc}$, are explicitly
presented. Thus, the Universe evolution at low redshifts determines the
angular distance $D_A$ in eq. (\ref{DA}), which depends on the initial data
on $H_0$ and $q_0$, while the sound horizon $r_s$ and redshift $z_d$ are
determined by fractions of baryonic and dark matters in the energy budget of
Universe. Since the evolution at low redshifts in the cosmological
extrapolation of MOND conforms with the supernovae data in the same way as it
was done in $\Lambda$CDM, the angular distances are almost coincident in both
these models, hence, the description of BAO data is determined by fitting the
sound horizon $r_s$. In the case of the cosmological extrapolation of MOND
all of non-relativistic matter is identified to baryons, i.e.
$\Omega_M=\Omega_b$. Hence, we find $\Omega_bh^2\approx 0.057$, i.e. it is
twice and a half greater than the baryonic fraction in $\Lambda$CDM. At such
the value of baryonic density we get the agreement with the BAO data (see
table  \ref{table}). In numerical estimations we have put $h=0.71$, $T=2.725$
K and $q_0=-0.775$, $z_t=0.375$ as it was in the previous section. Then,
$\Omega_b=0.113$. Such the doubling can point to necessity of accounting for
the baryonic matter outside luminous stars and visible gas, that is also
numerically consistent with the lack of visible baryonic matter in some
galactic clusters, which dynamics is described in MOND without any dark
matter (see \cite{rev-MOND1,rev-MOND2}).

At the same parameters we estimate the ``acoustic scale''
\begin{equation}\label{la}
    l_A=(1+z_*)\,\frac{\pi D_A(z_*)}{r_s(z_*)},
\end{equation}
which is measured in the spectra of CMBR temperature anisotropy by locations
of peaks of temperature fluctuations versus the multipole number in WMAP data
\cite{WMAP}. Here $z_*$ is the redshift of decoupling, when due to the
recombination of electrons with protons the medium becomes transparent for
photons (see analytical approximations for $z_*$ in terms of baryonic
density, matter density and Hubble constant in \cite{HS96}.). Calculating the
value of $D_A(z_*)$, we take into account for the contribution of
relativistic particles into the Hubble constant, since
$$
    \rho_\gamma+\rho_\nu=\rho_b\frac{1+z}{1+z_{eq}}.
$$
As we can see from table  \ref{table}, the cosmological extrapolation of MOND
is safely able to describe the large scale structure data given by the
measurements of both the baryon acoustic oscillations of visible matter and
the anisotropy of CMBR\footnote{We have not considered the ``shift
parameter'' defined in \cite{Bond:1997wr},
$\mathscr{R}=\sqrt{\Omega_MH_0^2}(1+z_*)D_A(z_*)/c$, since, as authors of
\cite{Bond:1997wr} have pointed out, this quantity can not be directly
extracted from the angular spectrum of CMBR anisotropy because of a poor
accuracy of measurements at low values of the multipole number, i.e. at large
angles of correlations, wherein the shift parameter makes a significant
influence on the amplitude of spectrum, while the high multipoles are not
sensitive to $\mathscr{R}$. It means that the choice of model for fitting the
spectrum in the region of acoustic oscillations at $l\sim l_A$ determines the
extrapolation to low values of multipoles, but it is not critical for the
data at low multipoles because of poor accuracy of data at low $l\ll l_A$.
Consequently, the value of shift parameter given in \cite{WMAP} is actually
the model dependent-extrapolation, indeed.}.

It is interesting that our argumentation about changing the MOND regime at
distances of the order of the scale for the spatial inhomogeneity of matter
distribution, in fact, is confirmed by the calculation of acoustic scale
$l_A$, which points to the angular size of inhomogeneity about
$360^\circ/l_A\approx 1.2^\circ$ in consistency with our initial assumptions.

\section{The Milgrom's acceleration}

Since the critical Milgrom's acceleration in MOND is experimentally measured
with the high accuracy in the gas-reach galaxies \cite{McGaugh}
\begin{equation}\label{go-tild-exp}
    \tilde g_0=(1.21\pm0.14)\cdot 10^{-10}
    {\mbox{\small m}}/{\mbox{\small s}^2},
\end{equation}
it makes sense to use this information in order to refine of parameters in
the cosmological extrapolation of MOND.

Indeed, within our approach, the Milgrom's acceleration is determined by the
angular scale of inhomogeneity in the matter distribution $\delta\theta$, so
that in Hubble units
\begin{equation}\label{g0t1}
    \frac{\tilde g_0}{cH_0}=K_0\delta\theta.
\end{equation}
However, the acoustic scale gives
\begin{equation}\label{angle}
    \delta \theta=\frac{2\pi}{l_A}.
\end{equation}
Then, by expressing $K_0$ in term of deceleration parameter $q_0$ and
redshift $z_t$ for the transition from the Universe deceleration to its
acceleration in accordance with (\ref{K0}), we find
\begin{equation}\label{g0t2}
    \frac{\tilde g_0}{cH_0}=\frac{4\pi}{l_A\Omega_b}\,\frac{q_0^2}{(1+z_t)^3-1},
\end{equation}
that allows us to plot the dependence of Milgrom's acceleration on the baryon
density at fixed values of $q_0$ and $z_t$, which are determined by the
supernovae data. Further, we can compare the result with the empirical value
of
\begin{equation}\label{got-empir}
    \frac{\tilde g_0}{cH_0}=0.180\pm0.022,
\end{equation}
known with the 12\%-accuracy after taking into account for the uncertainty of
the Hubble constant.
\begin{figure}[h]
\setlength{\unitlength}{1mm}
\begin{center}
  \begin{picture}(80,56)
  \put(1,1){
  \includegraphics[width=79\unitlength]{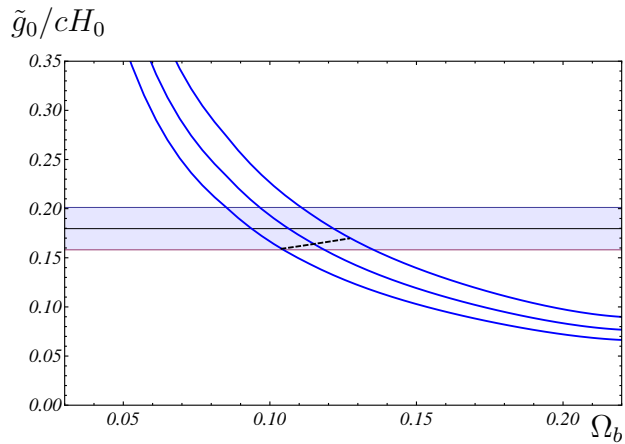}}
  \put(0,54){${\tilde g_0}/{cH_0}$}
  \put(77,0){$\Omega_b$}
  \end{picture}
\end{center}
  \caption{The Milgrom's acceleration ${\tilde g_0}/{cH_0}$ calculated
  in accordance to (\ref{g0t2}) at the deceleration parameter $q_0=-0.853$ and
  the redshift of zero acceleration $z_t=\{0.338,0.375,0.413\}$
  (the curves from top to bottom, respectively) as well as at given
  acoustic scale $l_A=302.6$ (the dashed line) versus the baryon fraction
  $\Omega_b$ in comparison with the experimental value (the shaded band).}
  \label{g0tilde}
\end{figure}

The comparison is shown in figure \ref{g0tilde}. It turns out that the
agreement of data on the supernovae (SN), baryonic acoustic oscillations
(BAO), acoustic scale (WMAP) and Hubble constant ($H_0$) with the Milgrom's
acceleration is reached at the deceleration parameter $q_0\approx-0.853$ and
accessible 10\%-variation of the transition redshift $z_t\approx0.375$. In
this way, the account for the small error in the value of acoustic scale
leads to
\begin{equation}\label{WMAP-BAO-H0-SN-g0}
    \begin{array}{rcl}
      \Omega_b &=& 0.115\pm0.012,  \\[1mm] \displaystyle
      \tilde g_0 &=& (1.10\pm0.03)\cdot 10^{-10}
    {\mbox{\small m}}/{\mbox{\small s}^2},\\[2mm]
    \frac{r_s(z_d)}{D_V(0.2)} &=& 0.185\pm 0.006,\\[2mm]
    \frac{r_s(z_d)}{D_V(0.35)} &=& 0.112\pm 0.004.
    \end{array}
\end{equation}

Thus, we successfully conform the logical consistency of two MOND regimes:
the cosmological limit and the locally inhomogeneous case; and we extract the
confident intervals for the Universe parameters.

\section{The evolution into the future}

\begin{figure}[th]
\setlength{\unitlength}{1mm}
  \begin{center}
\begin{picture}(79,48)
\put(1,0){
  \includegraphics[width=77\unitlength]{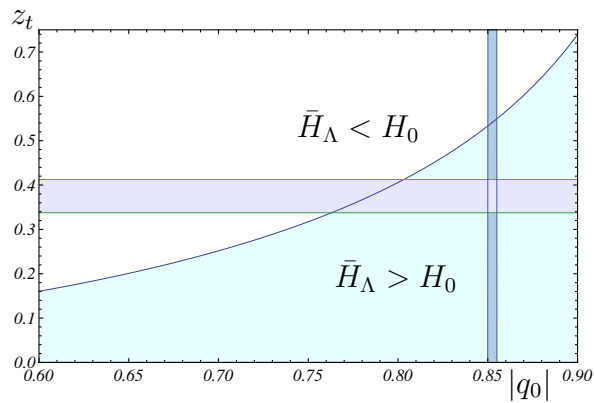}}
\put(68,-1){$|q_0|$}
\put(2,48){$z_t$}
\put(45,13){${\bar H}_\Lambda> H_0$}
\put(40,33){${\bar H}_\Lambda< H_0$}
\end{picture}
\end{center}
  \caption{Delimiting the future epochs versus the deceleration parameter
  $q_0$ and the redshift of zero acceleration $z_t$. The preferable fit of
  WMAP+BAO+SN+$H_0$+$\tilde g_0$ data is depicted by the crossing
  of the shaded bands.}\label{futur}
\end{figure}

In the future $a(t)\gg 1$, and the equation of cosmological MOND is reduced
to
\begin{equation}\label{future}
    \frac{\ddot a}{a}\,\frac{|\ddot a|}{a}=\frac12\,H_0^4K_0\Omega_b(1+z_t)^3,
\end{equation}
hence, according to (\ref{K0})
\begin{equation}\label{future2}
    \frac{\ddot a}{a}\,\frac{|\ddot a|}{a}=H_0^4q_0^2\,\frac{(1+z_t)^3}
    {(1+z_t)^3-1} ,
\end{equation}
so that the limit is the de Sitter Unverse with $a(t)=a_\star\exp[{\bar
H}_\Lambda (t-t_\star)]$, where the cosmological Hubble constant
\begin{equation}\label{bar-H}
    {\bar H}_\Lambda^4=H_0^4q_0^2\,\frac{(1+z_t)^3}
    {(1+z_t)^3-1} .
\end{equation}
Since $z_t>0$ and ${\bar H}_\Lambda\gtrless H_0$,
\begin{equation}\label{constr}
    q_0^2<1,\qquad
    z_t\lessgtr\frac{1}{\sqrt[3]{1-q_0^2}}-1.
\end{equation}
Therefore, the future epochs at ${\bar H}_\Lambda\gtrless H_0$ are
constrained by condition (\ref{constr}) as illustrated in figure \ref{futur}.

Substituting of $q_0=-0.853$ as obtained by fitting the data, into eq.
(\ref{constr}) gives
$$
    z_t\lessgtr0.54,
$$
so that according to our estimates for $z_t$ in the previous section, it
means that to the end of evolution ${\bar H}_\Lambda>H_0$ (see figure
\ref{futur}).

It is interesting to compare the cosmological term evaluated in MOND in terms
of ${\bar H}_\Lambda$ with its value in GR:
\begin{equation}\label{H-GR}
    H_\Lambda^2=\frac{8\pi G}{3}\rho_\Lambda=\frac12\,{\bar\Omega}_bH_0^2(1+z_t)^3,
\end{equation}
so that the equality of ${\bar H}_\Lambda=H_\Lambda$ takes place at
$$
    {\bar\Omega}_b^2=\frac{4q_0^2}{(1+z_t)^3}\,\frac{1}{(1+z_t)^3-1}.
$$
Numerically
$$
    \Omega_b\ll\bar {\Omega}_b.
$$
Therefore, the limit of de Sitter Universe and, hence, the problem of
small-scale cosmological constant cannot get an adequate description in GR,
while in order to consider the vacuum we need essentially to modify the
equation for the connection between the gravity and matter.

\section{Discussion and conclusion}

In this paper we have conformed the MOND to its cosmological extrapolation,
that allows us to eliminate hypothetical dark matter not only at scales of
spatial inhomogeneity of baryon matter distribution in galaxies and galactic
clusters, but also in the description of Universe evolution up to the
redshift of $z\approx 1.8$ in the observation of type Ia supernovae. In this
conformation, the angular size of inhomogeneity in the large scale structure
of Universe is essential.

Another question, related to the cosmology and modification of gravity, is
the necessity to postulate the dark matter in order to describe the CMBR
anisotropy in $\Lambda$CDM. To our point of view, this problem is also
closely connected to the usage of Friedmann model of expansion at redshifts
from zero to $z_*$, when the CMBR inhomogeneity was formed. Clearly,
parameters such as the contribution of cosmological term into the energy
budget of Universe and the fraction of dark matter can be mostly significant
at low redshifts, exactly when the modification of gravitational law becomes
remarkable at low accelerations of expansion. However, the forming of CMBR
essentially depends on the proportion between the baryons and dark matter. On
the other hand, the evolution of the baryonic density versus the redshift is
completely determined by the conservation law of  matter, hence, the
conformation of the baryon densities in both $\Lambda$CDM and MOND allows us
to generate similar conditions for the interaction between photons and
electrons, which density correlates with baryons. The problem of considering
the anisotropy evolution versus the redshift at the whole interval in the
framework of cosmological MOND expects a solution, i.e. the question about a
detailed investigation of CMBR anisotropy still remains open within the
cosmological extrapolation of MOND. In this respect we have to mention the
field theory-approaches in \cite{TVS} and \cite{Moffat}, which can, in
principle, provide us with complete calculations for the propagation of
perturbations, whereas the theory by J.Bekenstein \cite{TVS} includes the
MOND regime for inhomogeneous distribution of matter. However, we follow
another motivation based on the minimal extension of MOND that satisfies the
reasonable constraint on the transition to the cosmology of homogeneous
distribution of matter. In this way, we note that some integral quantities,
which values are deduced by the consideration of CMBR anisotropy, are derived
under the assumption of Friedmann evolution and they cannot be
straightforwardly transferred to the modified law of gravity. This note is
also valid as concerns for the observation of large scale structure, i.e. in
studying the baryon acoustic oscillations\footnote{Notice, in the framework
of $\Lambda$CDM the spatial inhomogeneity of both the matter and dark matter
are generated, say, by the primordial spectrum of inhomogeneity as given by
quantum fluctuations of a scalar filed having caused the Universe inflation
\cite{i-Guth,i-Linde,i-Albrecht+Steinhardt,i-Linde2,inflation}. Therefore,
the baryonic and dark matter initially has got common locations of
compression-decompression regions, that is consistent with the assumption
about the joint concentration of baryonic and dark matter. However, the
baryon oscillations having the electromagnetic nature, suggest that the
inhomogeneities of the photon-electron-baryon medium propagate as sound waves
decoupled from the dark matter (except the interaction due to gravitational
forces), while the inhomogeneities of dark matter do not propagate at all,
since there are no sound in the dark matter. During the evolution after the
baryon acoustic oscillations stopped, the gravitational interaction attracts
the centers of baryonic and dark matter concentration, that results in a
partial nearing the locations of matter lumps. The transfer function of
initial perturbations in the energy densities consists of two different parts
given by the terms of baryonic and dark matter. Consequently, initially
coinciding regions of concentration for the baryons and dark matter become
spatially separated, that creates different centers of gravitational
contraction of baryons and dark matter in the future evolution. Thus, the
existence of baryon acoustic oscillations challenges the dark matter
hypothesis by asking for a natural explanation for the coincident locations
of baryons and dark matter in galaxies, that rises the additional problem in
a dynamical derivation of the Tully--Fisher law in the framework of standard
cosmology.} (BAO) caused by the spatial inhomogeneity \cite{BAO}.
Nevertheless, we have analyzed the ``angular scale'' in the CMBR anisotropy
and parameters of baryon acoustic oscillations in the framework of
cosmological MOND and showed that MOND can confidently describe these
cosmological phenomena. Then, we can conclude that the cosmological MOND has
exhibited its positive potential for the adequate description of Universe
evolution at low redshifts as well as up to the redshifts of CMBR forming. We
have to note, of course, that the purpose of this paper has been not a
complex fitting the type Ia supernovae data, baryon acoustic oscillation and
CMBR anisotropy, but the aim has been the demonstration of opportunity to get
the successful application of cosmological MOND to those problems.

Finally, we have compared the critical acceleration calculated in the
framework of cosmological MOND with its empirical value and extract the
intervals of cosmological parameters, wherein there is the good agreement of
the model with observations of WMAP+BAO+SN+ $H_0$+$\tilde g_0$. In addition,
we have found that the modification of gravity law is essential in studying
the problem of cosmological constant.

Thus, the preliminary estimates performed in the framework of cosmological
MOND conceptually permit for the confidential description of Universe
evolution and point to the necessity to work out a cumbersome analysis of
complete base of the data on the supernovae, large scale structure,
baryogenesis and abundance of elements in the Universe etc., for the accurate
extraction of cosmological parameters, in particularly, the baryonic fraction
of energy. Nevertheless, in this paper we have shown that hypothetical dark
matter can be completely excluded from the cosmology by the appropriate
modification of gravity law at accelerations less than the critical value in
analogy to the case, when it was successfully done at the galactic scale in
MOND.

Notice, there are several papers, wherein the influence of modified law of
gravity to the Universe evolution at low accelerations was considered
\cite{Gao,ChLi2}, however, in those papers the acceleration is evaluated by
the temperature of apparent horizon $T_H=H/2\pi$ giving $g_H=2\pi T_H=H$,
while the critical acceleration is the constant value independent of
distance. In \cite{ChLi2} this approach follows the MOND at the linear
dependence of factor modifying the acceleration $\ddot a r$ on ratio
$g/g_H\ll 1$. Hence, it is clear that such the modification of gravity law at
the Hubble horizon is extrapolated inside the Hubble sphere, that can be
never conformed with our consideration.

\ack

This work was partially supported by the grant of Russian Foundations for
Basic Research 
10-02-00061, the grant of Special Federal Program ``Scientific and academics
personnel'' for the Scientific and Educational Center 2009-1.1-125-055-008,
and the work of T.S.A. was supported by the Russian President grant
MK-406.2010.2 as well as by the National Science Support Foundation program
``The best graduates of Russian Academy of Sciences''.

\end{document}